\documentclass[preprint,authoryear]{elsarticle}
\usepackage{comment}
\usepackage{pdflscape}
\usepackage{multirow}
\usepackage{graphicx}
\usepackage{xcolor}
\usepackage{caption}
\usepackage{subcaption}
\usepackage{hyperref}
\usepackage{rotating}

\usepackage{booktabs}
\usepackage{multirow}
\usepackage{lscape}

\newcommand{\gd}{$\mathrm{Gd}_2\left(\mathrm{SO}_4\right)_3\cdot8\mathrm{H}_2\mathrm{O}$}
\newcommand{\tbkgspec}{$\mathcal{B}$}
\newcommand{\tsampspec}{$\mathcal{S}$}
\newcommand{\mbkgspec}{\mathcal{B}}
\newcommand{\msampspec}{\mathcal{S}}

\journal{Applied Radiation and Isotopes}

\begin{document}

\begin{frontmatter}

\title{Background Shielding by Dense Samples in Low-Level Gamma Spectrometry}

\author[1]{M.~Thiesse\corref{cor1}}
\ead{m.thiesse@sheffield.ac.uk}

\cortext[cor1]{Corresponding author}
\address[1]{The University of Sheffield,\\Department of Physics and Astronomy, Sheffield, S3 7RH, United Kingdom}

\author[2]{P.~Scovell}

\address[2]{Boulby Underground Science Facility,\\Boulby Mine, Saltburn-by-the-Sea, Cleveland, TS13 4UZ, United Kingdom}

\author[1]{L.~Thompson}

\begin{abstract}
  In low activity gamma spectrometric measurements of large, dense samples, the bulk sample material shields the HPGe crystal from external background sources. If not accounted for in studies that utilise background-subtraction methods, this effect may result in systematic errors in the sample activity and detection limit estimation. We introduce a Monte Carlo based method to minimise the impact of this effect on sample gamma spectra. It is validated using simulated detector backgrounds and applied to a measurement of low-activity \gd. One main prerequisite for the correct application of this method is to know in advance the nuclides which contribute to the detector background spectrum and their spatial distribution. With a thorough understanding of the detector backgrounds, the method improves the accuracy of sensitive low-background measurements of low-activity samples. Even without knowing the background sources and their distribution, conservative results may still be presented that account for the potential systematic errors introduced by this background shielding effect.
\end{abstract}

\begin{keyword}
  gamma spectrometry \sep low-background \sep low-level activity \sep methods
\end{keyword}

\end{frontmatter}

\section{Introduction}
\label{sec:intro}

Gamma-ray spectrometry using high purity Germanium (HPGe) detectors is used to non-destructively analyse the radiopurity of material samples with low natural radioactivity \citep{hult2007low}. A deep understanding of the detector background radioactivity is necessary to effectively utilise this technology.

There are many sources of backgrounds, consisting of contributions from many sources around the detector and laboratory in general. For this paper, we distinguish two main categories: those ``external'' to the sample, and those ``internal''. The external sources are those which are farther from the HPGe crystal than the radial extent of the sample, such as:
\begin{enumerate}[--]
\item cosmogenic sources,
\item radioactivity in the laboratory environment and building materials,
\item $^{210}$Pb or other unstable isotopes in the shield,
\item decays of neutron-activated materials in the surrounding environment,
\item airborne Rn within the shield,
\item or radioactivity from parts of the sample container.
\end{enumerate}
Internal background sources originate from locations that are nearer to the HPGe crystal than the sample, such as:
\begin{enumerate}[--]
\item impurities in the HPGe crystal,
\item primordial nuclides in the crystal housing, window, screws, cold finger or other crystal mount components,
\item or radioactivity from the field-effect transistor (FET) or other front-end electronics.
\end{enumerate}

The community has expended much effort to identify, quantify and reduce the sources of background radiation in HPGe detectors. Even for detectors that were built to a high standard of radiopurity, by pre-screening construction materials~\citep{heusser2006,brodzinski1990}, operating deep underground~\citep{scovell2018}, using an active muon shield~\citep{povinec2004}, using ancient Pb with low $^{210}$Pb content~\citep{brodzinski1995}, purging Rn from within the shield~\citep{perez2022}, or other techniques~\citep{hult2013} \citep[See][for reviews of HPGe background mitigation techniques]{heusser1995,hult2007low}, the remaining background contribution from external and internal sources limits the HPGe detection limit ($L_D$)~\citep{hurtgen2000}.

Studies of underground low-background HPGe detectors showed that many of the backgrounds associated with $^{226}$Ra, $^{222}$Rn, $^{60}$Co and $^{40}$K were external to the sample, and backgrounds associated with $^{232}$Th progeny were mostly internal~\citep{brodzinski1995,hult2008mercury}. Shields comprised of low-background Pb with Cu inner lining are typical for modern HPGe detector systems~\citep[e.g.][]{scovell2018}, though materials can vary across manufacturers and depend on budget and date of construction.

Effective use of HPGe technology to achieve unambiguous detection of low sample activities requires a thorough characterisation of the detector's response to background radiation. To calculate a sample activity there are several physical processes or measurements to consider when characterising a detector, including the full-energy peak efficiency of detection, self-attenuation of photons by dense materials, temporal background variations, coincidence summing, peak-correlated background variations~\citep{berlizov2007}, and the shielding of background radiation by the sample~\citep{hult2007low}.

There are many published HPGe measurements of samples with ultra-low activities on the mBq/kg or $\mu$Bq/kg scale. Construction of next-generation rare-event physics experiments, such as neutrinoless double-beta decay~\citep{maneschg2008}, cold dark matter~\citep{akerib2020lux} and supernova relic neutrinos~\citep{marti2020}, frequently require such ultra-low activity materials to reduce backgrounds and improve sensitivity. Achieving such low limits on the material activity requires careful consideration of all potential systematic errors, HPGe detector calibrations and sources of laboratory backgrounds.

The shielding of background radiation by a large, dense, radiopure sample can cause a significant underestimate of the sample activity and an overestimate of $L_D$ when the nuclides measured in the sample also contribute to the background~\citep{bastos2008}. An unphysical negative background-subtracted net count rate~\citep{hult2008mercury} beyond what can be reasonably explained by stochastic fluctuation indicates this effect in a sample spectrum analysis. So far, there is no published method for correcting this effect in an HPGe detector measurement. We will show that a Monte Carlo (MC) based correction can be implemented to eliminate the statistically significant erroneous net count rates caused by background attenuation by the dense sample material, thereby allowing more accurate measurements of sample activities near to or lower than the detector background level.

\section{Problem Statement}
\label{sec:prob}

In a typical HPGe measurement, two energy spectra are measured: the detector background, \tbkgspec, and the sample spectrum, \tsampspec. Given a particular experimental condition, \tbkgspec\ consists of the background contribution from all sources apart from the sample and might include a reference standard. Under the same experimental conditions, but replacing the standard with the sample, \tsampspec\ consists of all contributions from \tbkgspec\ and the sample activity, less the activity of the standard.

The sample activity is reported relative to the activity of the standard. An absolute nuclide activity measurement requires a standard with zero radioactivity from that nuclide. When the measured nuclides in the sample are also present in the background or standard material, especially for naturally occurring radioactive materials (NORM), making a high-density radiopure standard is usually impractical. Instead, a close approximation is generally achieved by measuring \tbkgspec\ with a Rn-free gas in the place of the sample, which effectively becomes the standard.

By measuring the background without a standard of equal density and composition to the sample, the attenuation of external photons changes the background contribution to both \tbkgspec\ and \tsampspec. Since Compton scattering depends on the material electron density, internal backgrounds will also interact differently with the sample and standard, altering the backscattered contribution to \tsampspec\ compared with \tbkgspec.

\begin{figure}
  \centering
  \includegraphics[width=0.5\textwidth]{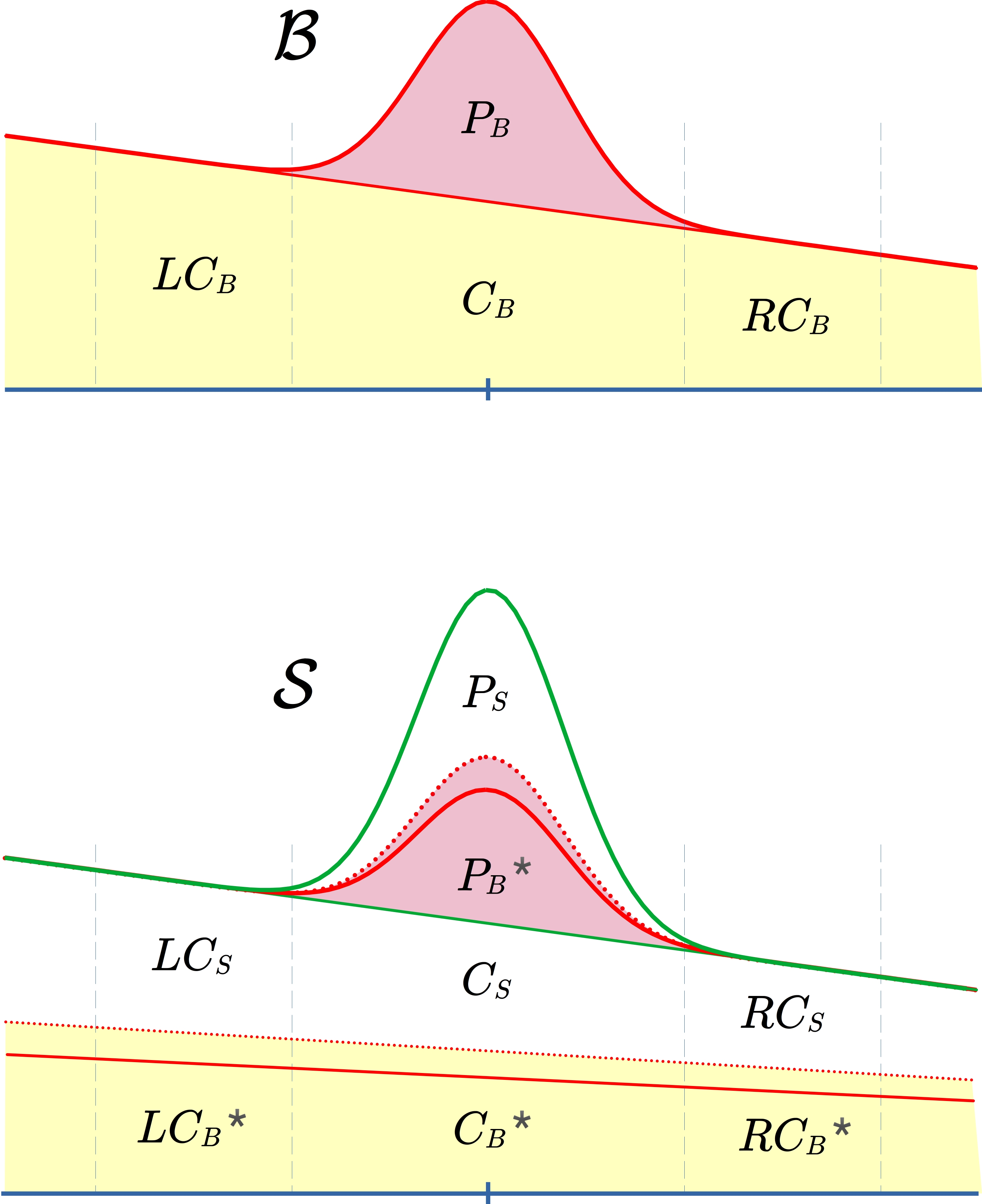}
  \caption{Components of measured background, $\mathcal{B}$, and sample, $\mathcal{S}$, spectra for a particular energy. Subscripts denote the source of the counts in each spectrum; $B$ if the counts were caused by photons that originated in the detector background, and $S$ for the counts associated with photons emitted by the sample. Asterisks denote the effect of the sample material in altering the background contribution to \tsampspec.}
  \label{fig:diag}
\end{figure}

To illustrate this, we consider the spectrum components in \tbkgspec\ and \tsampspec, as in Figure~\ref{fig:diag}. In this case, the nuclide under measurement is present in the sample and the background. Whereas the full-energy ($E=E_\gamma$) peak contribution to \tbkgspec\ is $P_B$, the peak background contribution to \tsampspec\ is affected by some amount due to differences in attenuation of the sample compared with the standard, $P_B^*$. A similar effect is expected where the Compton continuum contribution from background sources to \tbkgspec, $C_B$, is altered in \tsampspec\ due to different matter effects in the sample measurement, $C_B^*$.

The fundamental measurable values in either spectrum are the underlying background continuum and net (background-subtracted peak) count rates in the region of interest (ROI) of a particular $E_\gamma$. Depending on the spectrum count rate and if a peak is observed, there are many methods for estimating the net peak area and interpolating the underlying Compton continuum within the ROI. Even for low counts where Poissonian statistics dominate, the underlying measurable quantities are still relevant. For \tbkgspec, 
\begin{equation} \label{eq:netbkg}
  \mbkgspec\left(E_\gamma\right)_{\mathrm{net}}=P_B
\end{equation}
and
\begin{equation} \label{eq:bkgbkg}
  \mbkgspec\left(E_\gamma\right)_{\mathrm{bkg}}=C_B.
\end{equation}
For \tsampspec,
\begin{equation} \label{eq:netsamp}
  \msampspec\left(E_\gamma\right)_{\mathrm{net}}=P_S+P_B^*
\end{equation}
and
\begin{equation} \label{eq:bkgsamp}
  \msampspec\left(E_\gamma\right)_{\mathrm{bkg}}=C_S+C_B^*.
\end{equation}
There are no common terms between these estimates which would allow a calculation of the activity contribution of the sample, $P_S$. Additional information is needed about the material effects of the sample and standard on the background sources and the relative contributions of internal and external sources to \tbkgspec\ to perform an unbiased activity analysis.

The significance of the background shielding effect depends on many factors like the difference in densities of the sample and standard, the solid angle coverage of the sample and standard around the HPGe crystal, the relative activity difference between the background and sample and the relative contributions of internal and external background sources to \tbkgspec. Consider an experiment in which the detector background is measured with a Rn-free gas standard and the activity of a high-density sample is expected to be low (for example, ultra-pure stainless steel, OFHC copper, mercury, PTFE, or \gd). If the external background nuclide activity is comparable to or higher than the sample nuclide activity, the attenuation of background photons by the sample material can significantly alter the background contribution to \tsampspec\ compared with \tbkgspec. If the sample covers a large solid angle of the HPGe crystal, such as a Marinelli beaker, the effect is amplified as more background photons will experience attenuation in the sample.

In this representative example, the sample will attenuate the background component, $P_B^*<P_B$. If the background contains a measurable activity of the particular nuclide and this attenuation is not accounted for, then following background subtraction from \tsampspec, the sample activity will be underestimated. Where the activity contribution of the sample to \tsampspec\ is less than $P_B-P_B^*$, an estimate of the sample activity may appear significantly negative. Such statistically significant negative count rates are observed in the literature~\citep{hult2008mercury} and measurements of ultrapure \gd\ at Boulby Underground Laboratory (see Section~\ref{sec:realmeas}).

Even if the background does not have any contribution from the same nuclide that is under measurement in the sample, the contribution of the background Compton continuum at $E_\gamma$ from all other background nuclides will be affected by the sample material. Even if $P_B=P_B^*=0$ for some nuclides, it is still likely that $C_B \neq C_B^*$.

Analyses must consider the effect of background attenuation by large, dense samples to reduce detection limits in low-background HPGe detector systems. We will describe an MC-based method for calculating the relationship between $P_B$ and $P_B^*$ and between $C_B$ and $C_B^*$ for a given sample material, geometry and background source distribution.

By correcting for the material effects of the sample on the detector background, there remains no benefit to measuring the detector background with a standard of equal density to the sample since the background attenuation can be modelled using verified and tested MC codes. For the remainder of this discussion, we assume that \tbkgspec\ is measured with a Rn-free gas standard. The measured sample will have a higher density than the standard, so will more strongly attenuate background photons.

\section{Proposed Method of Correction}
\label{sec:correcting}

The intensity of a beam of photons travelling through matter is proportional to the initial beam intensity, where the proportionality factor depends on the photon energy, material electron density, and photon path length. Because the locations and intensities of all background sources in an HPGe measurement are assumed to be constant over time, we may consider the surviving intensity of external background photons at the HPGe in the background measurement to be proportional to the surviving intensity in the sample measurement. Therefore, $P_B$ is proportional to $P_B^*$,
\begin{equation} \label{eq:corrpeak}
  P_B^* = a \cdot P_B.
\end{equation}
The empirical proportionality factor, $a$, is a function of the material properties of the sample and standard, the photon energy, $E_\gamma$, the geometry of the shield, sample and detector system and the background source distribution. Since the sample density is assumed to be greater than the density of the Rn-free gas standard, $a \leq 1$.

The relationship between the full-energy peak count rates in \tbkgspec\ and \tsampspec\ is defined by combining Equations~\ref{eq:netbkg}, \ref{eq:netsamp}, and \ref{eq:corrpeak},
\begin{equation} \label{eq:netsampcorr}
  P_{S} = \msampspec\left(E_\gamma\right)_{\mathrm{net}} - a \cdot \mbkgspec\left(E_\gamma\right)_{\mathrm{net}}.
\end{equation}

There are many ways the Compton continuum might change when a sample is placed on a detector. Photon scattering cross-section differences between the sample and standard may increase or decrease the Compton continuum of \tsampspec\ relative to the Compton continuum of \tbkgspec. For instance, background photons that, on average, deposit their full energy in the HPGe crystal may experience additional scattering within the sample material, causing the full-energy peak to decrease and portions of the Compton continuum for $E<E_\gamma$ to increase. It is especially apparent in the case of backscattering for photons with $E_\gamma<500$~keV. Even the scattered photons, on average, will be more attenuated by the sample. Despite the many ways the placement of the sample on the detector may change the background continuum of \tsampspec, all possibilities are accounted for by introducing another empirical proportionality factor, $b$, to relate the Compton continuum contribution of background sources between both spectra. The background continuum contribution in \tsampspec\ is altered to include these effects,
\begin{equation} \label{eq:corrcontinuum}
  C_{B}^*=b\cdot C_{B}
\end{equation}
where $b$ is a function of $E_\gamma$, the material properties, the measurement geometry and the background source distribution.

The total background count rate within the ROI, $B$, is required to calculate $L_D$. However, due to the attenuation of background photons by the sample, the total background count rate that is relevant to the sample spectrum is adjusted according to the factors $a$ and $b$ and written in terms of the measurable spectrum quantities in Equations~\ref{eq:netbkg} and \ref{eq:bkgbkg}:
\begin{equation} \label{eq:totbkgcorr}
  B = P_B^*+C_B^* = a\cdot \mbkgspec\left(E_\gamma\right)_{\mathrm{net}} + b\cdot\mbkgspec\left(E_\gamma\right)_{\mathrm{bkg}}.
\end{equation}
This corrected definition of the background count rate constitutes an unbiased estimate of the true detector sensitivity to the sample activity, even if the nuclide under measurement does not contribute to \tbkgspec.

The correction, $a$, is the factor by which the full energy peak area changes due to background attenuation by the sample. The factor $b$ is the ratio of the Compton continuum in \tbkgspec\ to the sample-altered background contribution to the continuum in \tsampspec. While $a$ and $b$ would be directly calculable from Equations~\ref{eq:corrpeak} and \ref{eq:corrcontinuum} by using a radiopure standard of the same material and geometrical properties of the sample, it is impractical to create such a standard. Moreover, since the mechanisms in which the underlying Compton continuum beneath a full-energy peak change are diverse and plentiful, calculating an analytical expression for $b$ is not straightforward. Therefore, MC simulation is employed to overcome these difficulties. Since laboratories typically already employ validated MC methods for estimating detector properties such as detection efficiencies and true coincidence summing factors, they can be used to correct for the background shielding effect without significant further effort. The main prerequisite for calculating $a$ and $b$ is a detailed understanding of the background source distribution for the detector and shield system. 

\subsection{A Simple Background Model}
\label{sec:simplebkg}

Modelling the HPGe background is a time-consuming procedure with some recent success in the literature~\citep{medhat2014,breier2018}. Often, detector background sources are varied and of unknown origin or relative activity. However, if a realistic model of the detector background can be simulated, the factors $a$ and $b$ are straightforward to estimate.

For this paper, we consider a basic model of the detector background assuming an isotropic distribution of sources around the HPGe crystal. As a typical low-background HPGe detector shield includes full angular coverage of low-background Pb and typically one or more layers of X-ray blocking material such as Cu or Cd, it is a reasonable simplification to simulate the external backgrounds originating from the surface of a spherical shell centred around the HPGe crystal centre-of-mass and with a radius large enough to contain all of the sample and the detector. The photons are emitted isotropically from their starting locations evenly spaced on the shell surface and with a uniform energy spectrum up to 3~MeV. On the other extreme, an internal background model is devised where photons are emitted with the same uniform energy distribution and isotropic directionality, but from the outer surface of the HPGe crystal housing.

These simplified background models are simulated twice: with the sample and without. The sample and Rn-free gas standard are simulated with no inherent activity so that all energy deposits detected in the HPGe crystal originated from background sources, allowing $P_S=C_S=0$. The resulting background detection efficiency and the total deposited energy spectrum are recorded.

Because the detector, shield, and sample geometries are taken into account already in the particle transport code of the MC simulation, the resulting background detection efficiency of \tbkgspec, $\varepsilon_{B}\left(E_{\gamma}\right)$, and the background detection efficiency of \tsampspec, $\varepsilon_{B}^*\left(E_{\gamma}\right)$, represent the cumulative expected change in the background contribution due to the sample and the background given the system geometry. Since the efficiency is proportional to the count rate, the ratio of efficiencies represents $a$,
\begin{equation} \label{eq:asim}
  a\left(E_\gamma\right)=\frac{\varepsilon_{B}^*\left(E_{\gamma}\right)}{\varepsilon_{B}\left(E_{\gamma}\right)}.
\end{equation}
The simulated background spectrum, $\mbkgspec\left(E_\gamma\right)$, and the background spectrum with the sample, $\mathcal{B}^*\left(E_\gamma\right)$, are divided to describe the change in the Compton continuum as a function of $E_\gamma$:
\begin{equation} \label{eq:bsim}
  b\left(E_\gamma\right)=\frac{\mathcal{B}^*\left(E_{\gamma}\right)}{\mathcal{B}\left(E_{\gamma}\right)}.
\end{equation}

In reality, background photons come from internal and external sources. An ideal estimate of $a$ and $b$ should account for the distribution of nuclide source locations and relative contributions to \tbkgspec. We can derive an expression for both correction factors using the simplified background components. The net peak contribution to \tbkgspec\ is comprised of internal (I) and external (E) contributions,
\begin{equation} \label{eq:netbkg-intext}
  \mbkgspec\left(E_\gamma\right)_{\mathrm{net}}=P_B=P_I+P_E.
\end{equation}
The Compton continuum can also be expressed using the background source components,
\begin{equation} \label{eq:bkgbkg-intext}
  \mbkgspec\left(E_\gamma\right)_{\mathrm{bkg}}=C_B=C_I+C_E.
\end{equation}
The internal and external backgrounds contribute independently to \tbkgspec\ and are altered independently by the sample. Therefore, we can expand the net peak contribution to \tsampspec,
\begin{equation} \label{eq:netsamp-intext}
  \msampspec\left(E_\gamma\right)_{\mathrm{net}}=P_S+P_B^*=P_S+P_I^*+P_E^*,
\end{equation}
and the Compton continuum contribution to \tsampspec,
\begin{equation} \label{eq:bkgsamp-intext}
  \msampspec\left(E_\gamma\right)_{\mathrm{bkg}}=C_S+C_B^*=C_S+C_I^*+C_E^*.
\end{equation}

Let us define the fractions of the background contributions from internal and external sources to be $\alpha$ and $\beta$, respectively. Defining $\alpha+\beta=1$ and using Equations~\ref{eq:corrpeak} and \ref{eq:corrcontinuum}, we can express Equation~\ref{eq:netsampcorr} in terms of the combination of internal and external background contributions:
\begin{equation} \label{eq:netsampcorr-intext}
  P_S=\msampspec\left(E_\gamma\right)_{\mathrm{net}}-\left(\alpha a_I + \left(1-\alpha\right)a_E\right)\cdot\mbkgspec\left(E_\gamma\right)_{\mathrm{net}}.
\end{equation}
By equivalence, we see that the correction factor, $a$, related to the total background contribution is the weighted arithmetic mean of the internal- and external-only correction factors,
\begin{equation} \label{eq:combineda}
  a=\alpha a_I+\left(1-\alpha\right)a_E.
\end{equation}
Each background nuclide component generally will have a different value for $\alpha$.

The total background contribution in the ROI which is relevant to the sample measurement, $B$, can also be expressed in terms of the internal and external correction factors, $a$ and $b$. Using Equations~\ref{eq:corrpeak} and \ref{eq:corrcontinuum}, Equation~\ref{eq:totbkgcorr} becomes
\begin{equation} \label{eq:totbkgcorr-intext}
  B = \left(\alpha a_I+\left(1-\alpha\right)a_E\right)\cdot \mbkgspec\left(E_\gamma\right)_{\mathrm{net}} + \left(\alpha b_I+\left(1-\alpha\right)b_E\right)\cdot \mbkgspec\left(E_\gamma\right)_{\mathrm{bkg}}.
\end{equation}
Analogous to Equation~\ref{eq:combineda}, $b$ is the weighted arithmetic mean of the internal- and external-only correction factors,
\begin{equation} \label{eq:combinedb}
  b=\alpha b_I+\left(1-\alpha\right)b_E.
\end{equation}

If a particular background source is only internal, the associated $E_\gamma$ peak areas will not be affected ($a_I = 1$), but the Compton continuum would show an enhanced ($b_I \geq 1$) backscattered contribution in \tsampspec. For external backgrounds such as $^{210}$Pb, $^{222}$Rn, $^{40}$K, $^{60}$Co, or other radionuclides present in the shield materials, the peak areas would either remain unchanged or attenuated by the sample depending if the photons pass through it ($a_E \leq 1$). The sample may enhance or attenuate the cumulative continuum contribution from all external sources ($b_E>0$). In all of these cases, the cumulative effect of the sample on the peak areas and continua for all background sources depends on the particular material and geometrical properties of the sample and the location of background sources. Table~\ref{tab:ab} summarises the allowable values for $a$ and $b$ for internal and external sources when the sample has a higher attenuation coefficient than the standard.

\begin{table}[tbp]
\centering
\begin{tabular}{@{}ccc@{}}
\toprule
\begin{tabular}[c]{@{}c@{}}Internal\\ Backgrounds\end{tabular} &
  \begin{tabular}[c]{@{}c@{}}External\\ Backgrounds\end{tabular} &
  \begin{tabular}[c]{@{}c@{}}All\\ Backgrounds\end{tabular} \\ \midrule
$a_I=1$ &
  $a_E\leq 1$ &
  $a\leq 1$ \\ \midrule
$b_I\geq 1$ &
  $b_E > 0$ &
  $b>0$ \\ \bottomrule
\end{tabular}
\caption{The allowable values for $a$ and $b$ for both types of background sources where the sample is more dense than the standard.}
\label{tab:ab}
\end{table}

\section{Validation of Correction Factors With An Example}
\label{sec:exsim}

To validate this definition of the correction factors $a_{I,E}$ and $b_{I,E}$, we consider a simulation of a real sample and detector with the simplified background model. The procedure set out in Section~\ref{sec:correcting} is used to determine the correction factors, then they are implemented into the analysis of a simulated sample activity measurement.

\subsection{MC Simulation}
\label{sec:mcsim}

A GEANT4~\citep{geant4} simulation of a 2.15~kg p-type coaxial HPGe detector is considered along with a sample of 5~kg of \gd\ packed in a Marinelli beaker (448G-E from GA-MA \& Associates, Inc.). The \gd\ has a bulk packed powder density of $1.84\pm0.11$~g/cm$^3$, where the error reflects the uncertainty in material fill height and sample mass measurement. The sample is initially assumed in the simulation to be free from all inherent radioactivity to assess its background attenuating properties.

\begin{figure}
  \begin{subfigure}{0.49\textwidth}
    \centering
    \includegraphics[width=\textwidth]{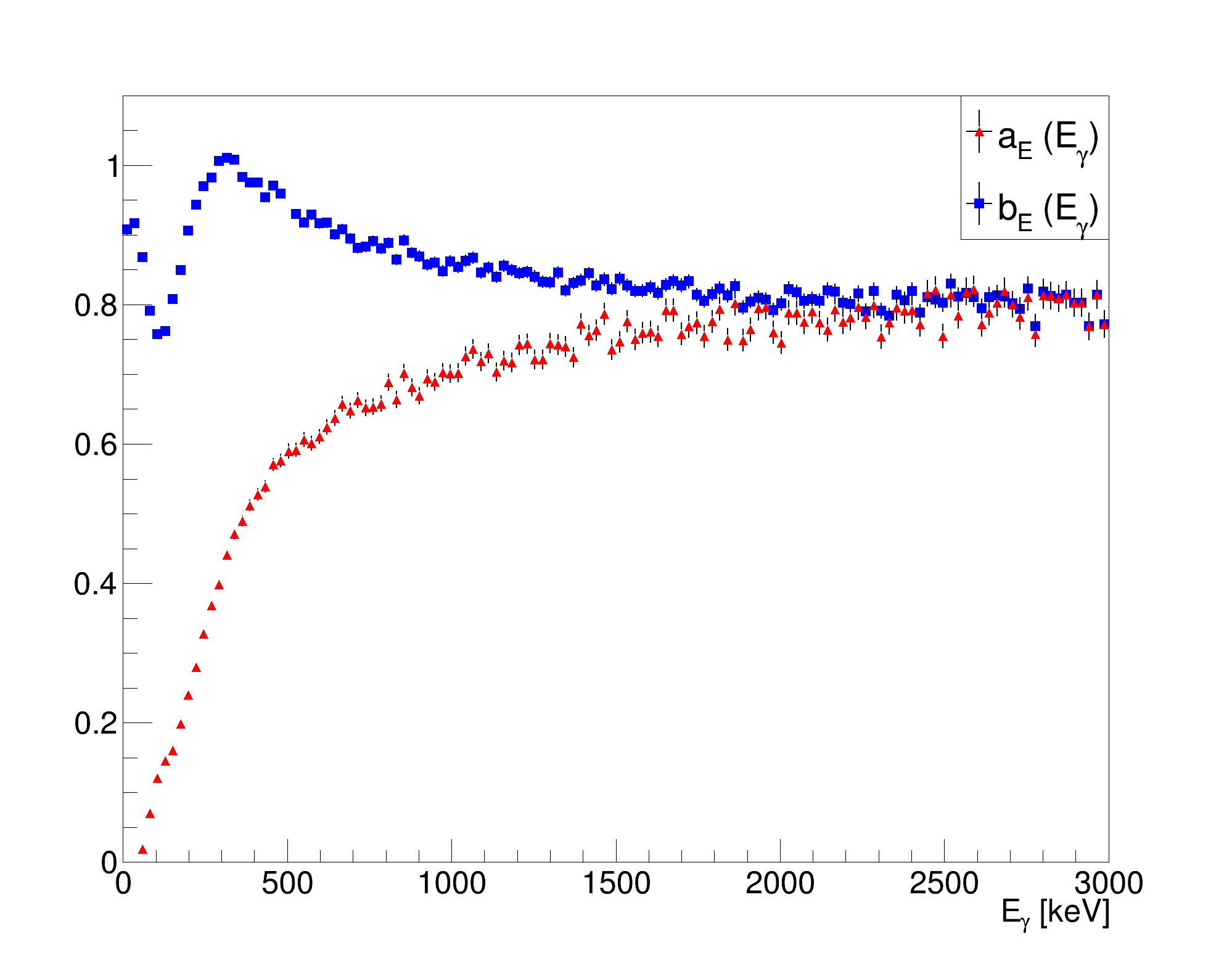}
    \caption{\label{fig:externalab}}
  \end{subfigure}
  \hfill
  \begin{subfigure}{0.49\textwidth}
    \centering
    \includegraphics[width=\textwidth]{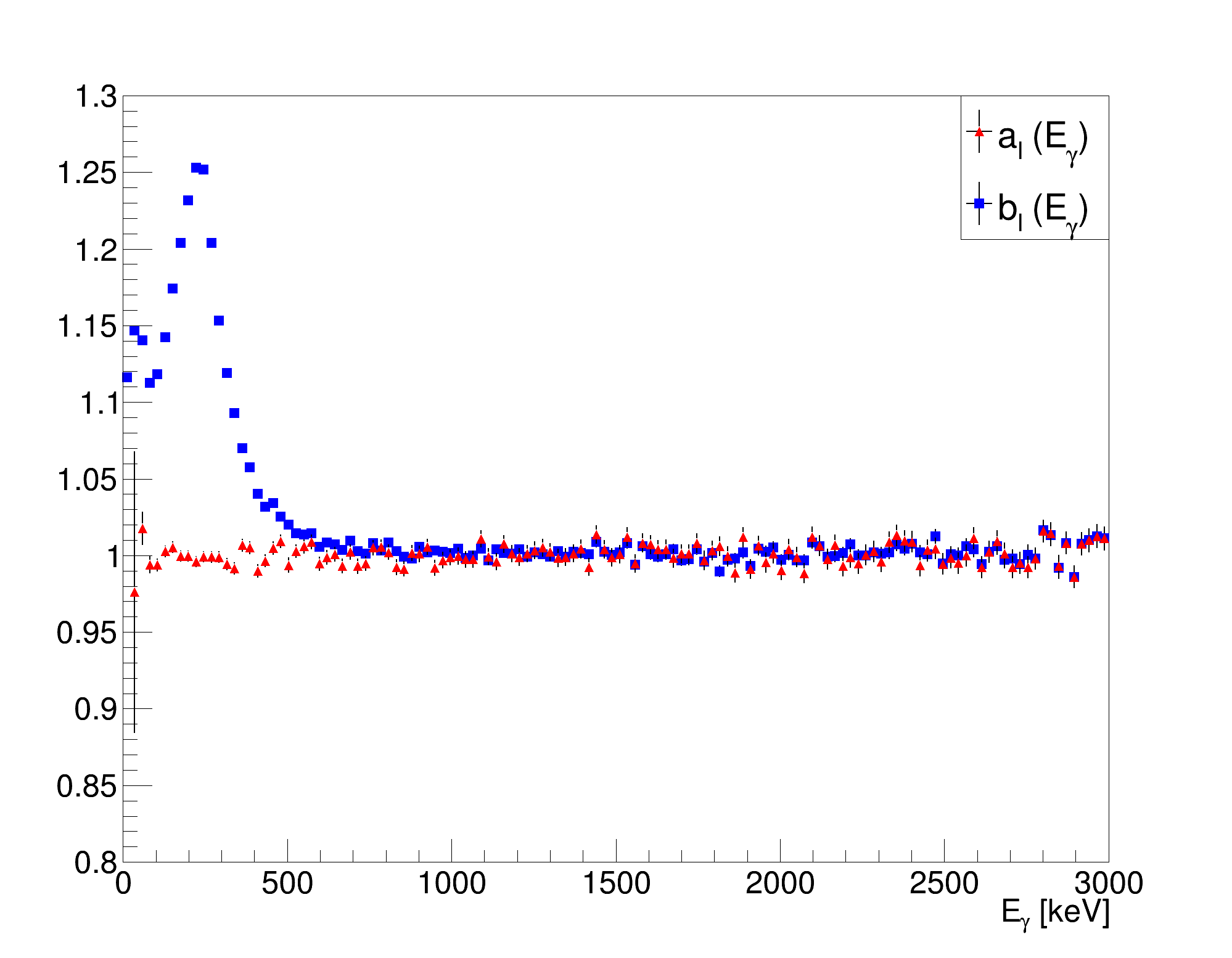}
    \caption{\label{fig:internalab}}
  \end{subfigure}
  \caption{Simulated correction factors, $a$ and $b$, for (a) external and (b) internal background sources for a sample of \gd\ in a Marinelli beaker.}
  \label{fig:corrfactors}
\end{figure}

The simplified internal and external models of detector background are simulated as in Section~\ref{sec:correcting} for the detector with the sample and with the \gd\ replaced with air. The full-energy background efficiency is determined in all cases, then the with- and without-sample efficiencies are divided to get $a_E\left(E_\gamma\right)$ and $a_I\left(E_\gamma\right)$ (Figure~\ref{fig:corrfactors}). The total deposited energy spectra with and without the sample are divided to get $b_E\left(E_\gamma\right)$ and $b_I\left(E_\gamma\right)$ (Figure~\ref{fig:corrfactors}). Note that the simulated $a_{I,E}$ and $b_{I,E}$ values agree with the predicted allowable values in Table~\ref{tab:ab}.

\begin{figure}
  \begin{subfigure}{0.49\textwidth}
    \centering
    \includegraphics[width=\textwidth]{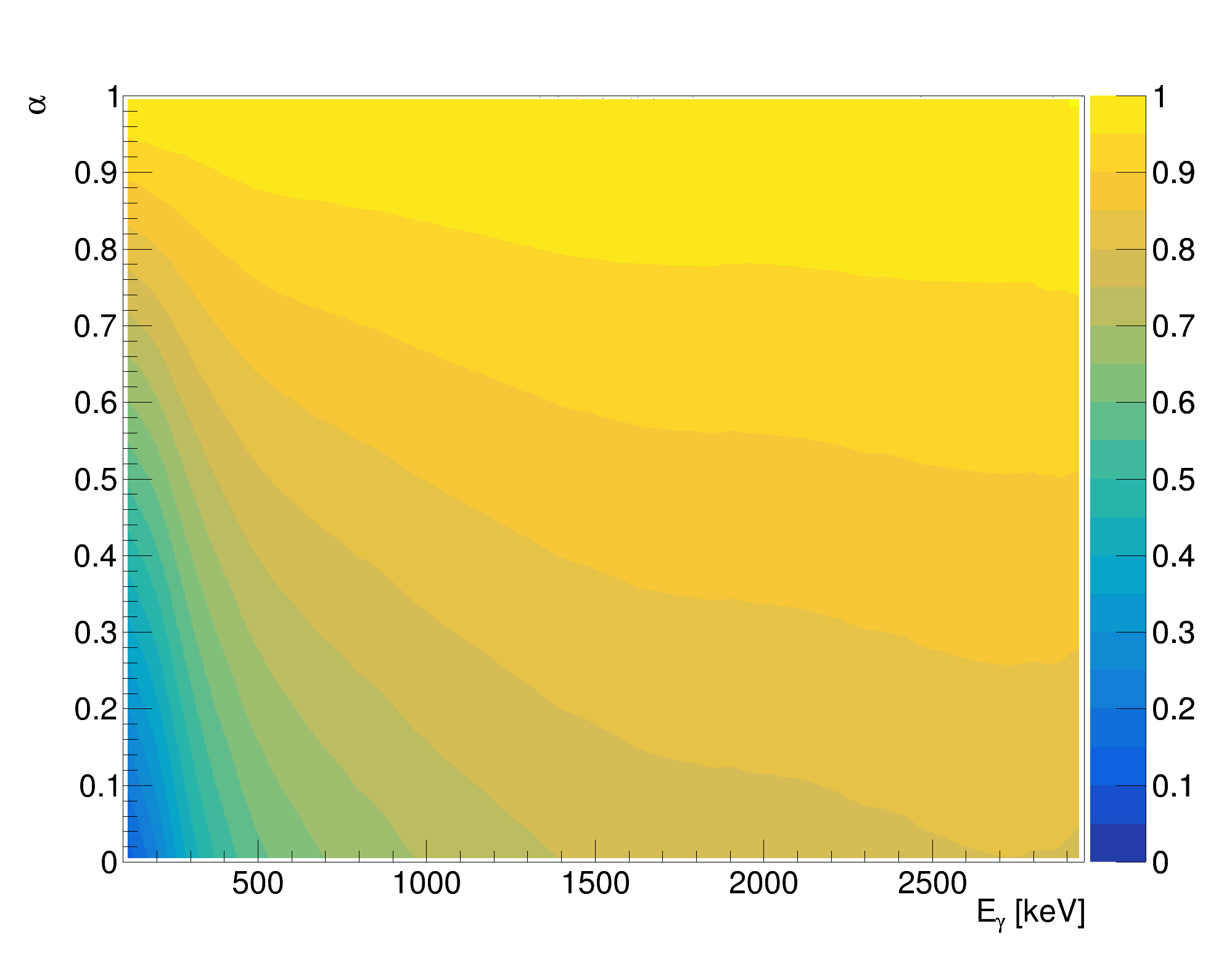}
    \caption{\label{fig:a-alpha}}
  \end{subfigure}
  \hfill
  \begin{subfigure}{0.49\textwidth}
    \centering
    \includegraphics[width=\textwidth]{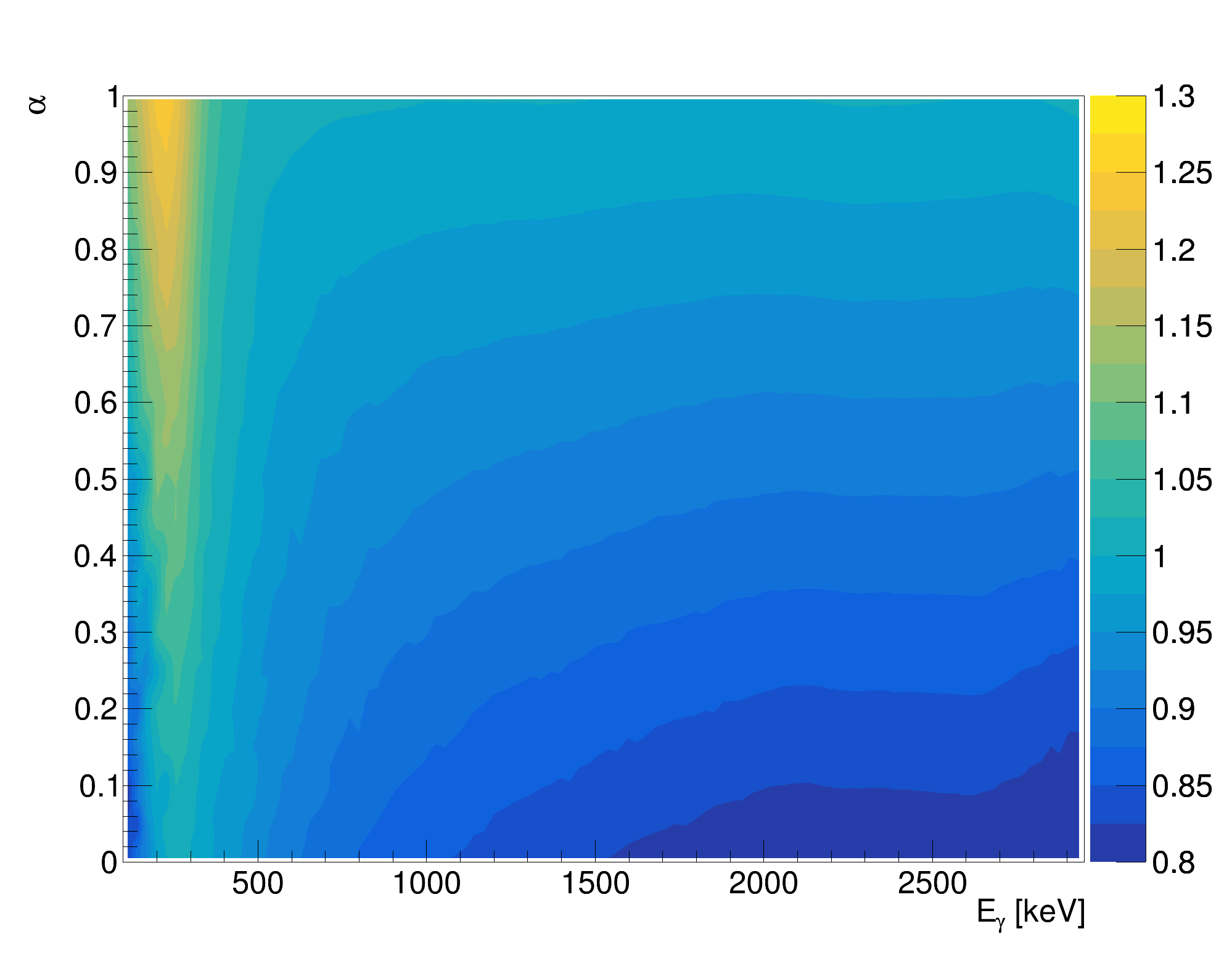}
    \caption{\label{fig:b-alpha}}
  \end{subfigure}
  \caption{Simulated correction factors (a) $a$ and (b) $b$ for a Marinelli beaker sample geometry, as functions of the fractional contribution of internal backgrounds to \tbkgspec, $\alpha$.}
  \label{fig:ab-alpha}
\end{figure}

In the current simplified model where backgrounds are isotropic around the HPGe crystal, the correction factors, $a$ and $b$, for this particular sample are shown as a function of photon energy and the fractional influence of internal ($\alpha=1$) and external ($\alpha=0$) background sources to \tbkgspec\ (Figure~\ref{fig:ab-alpha}).

\subsection{Validation of Correction Factors}
\label{sec:impl}

The simulated correction factors, $a$ and $b$, were implemented into the analysis procedure according to Equations~\ref{eq:netsampcorr-intext} and \ref{eq:totbkgcorr-intext} then validated using a simulated Marinelli beaker sample activity measurement with semi-realistic NORM backgrounds. The energy distribution of NORM backgrounds in each simulation is fixed and includes $^{238}$U, $^{232}$Th, $^{235}$U, $^{40}$K, $^{60}$Co and $^{137}$Cs at levels typical for some deep underground HPGe detectors~\citep{scovell2018}. The spatial distribution of internal and external backgrounds is isotropic, as in Section~\ref{sec:correcting}.

The activity of the sample is near the true $L_D$ for the simulated HPGe detector. For each of $^{238}$U, $^{235}$U, $^{232}$Th, $^{40}$K, and $^{176}$Lu, 0.5 mBq/kg is simulated, and 0.1 mBq/kg of $^{138}$La. To study the sensitivity of the detector to low sample activities in the presence of varying levels of background, a variable proportion of photons is simulated from the background relative to the sample emission. 

For each level of background activity, 100 repeated simulations are performed to eliminate the stochastic error. For the following discussion, the means and standard deviations of measured activities and $L_D$s are reported. The reported $L_D$ is calculated according to~\citet{hurtgen2000}, 
\begin{equation}\label{eq:LD}
  L_D=u^2+u\sqrt{8 B + 8 + u^2},
\end{equation}
with a one-tailed coverage factor of $u=1.645$ to give a 95\% confidence level.

\subsubsection{External Backgrounds}
\label{sec:externalbkg}

For purely external backgrounds ($\alpha=0$), the effect of background shielding by the sample is maximal. The 609.3~keV characteristic photons from $^{214}$Bi and the 295.2~keV photons from $^{214}$Pb are usually critical for analysing the activity of late-chain $^{238}$U in a sample, and they experience different levels of background shielding corrections (Figure~\ref{fig:externalab}).

\begin{figure}
  \begin{subfigure}{0.49\textwidth}
    \centering
    \includegraphics[width=\textwidth]{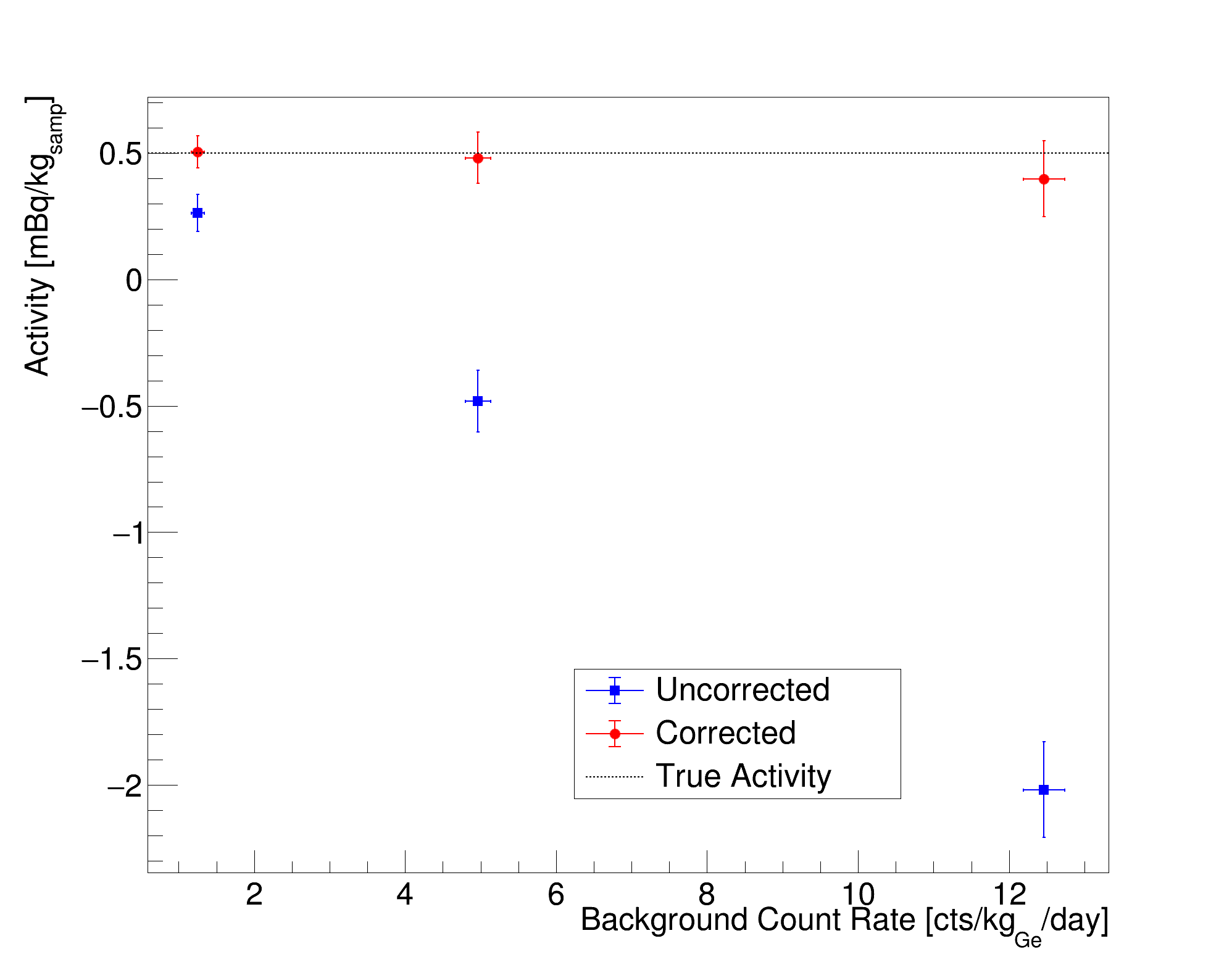}
    \caption{\label{fig:activity_609_external}}
  \end{subfigure}
  \hfill
  \begin{subfigure}{0.49\textwidth}
    \centering
    \includegraphics[width=\textwidth]{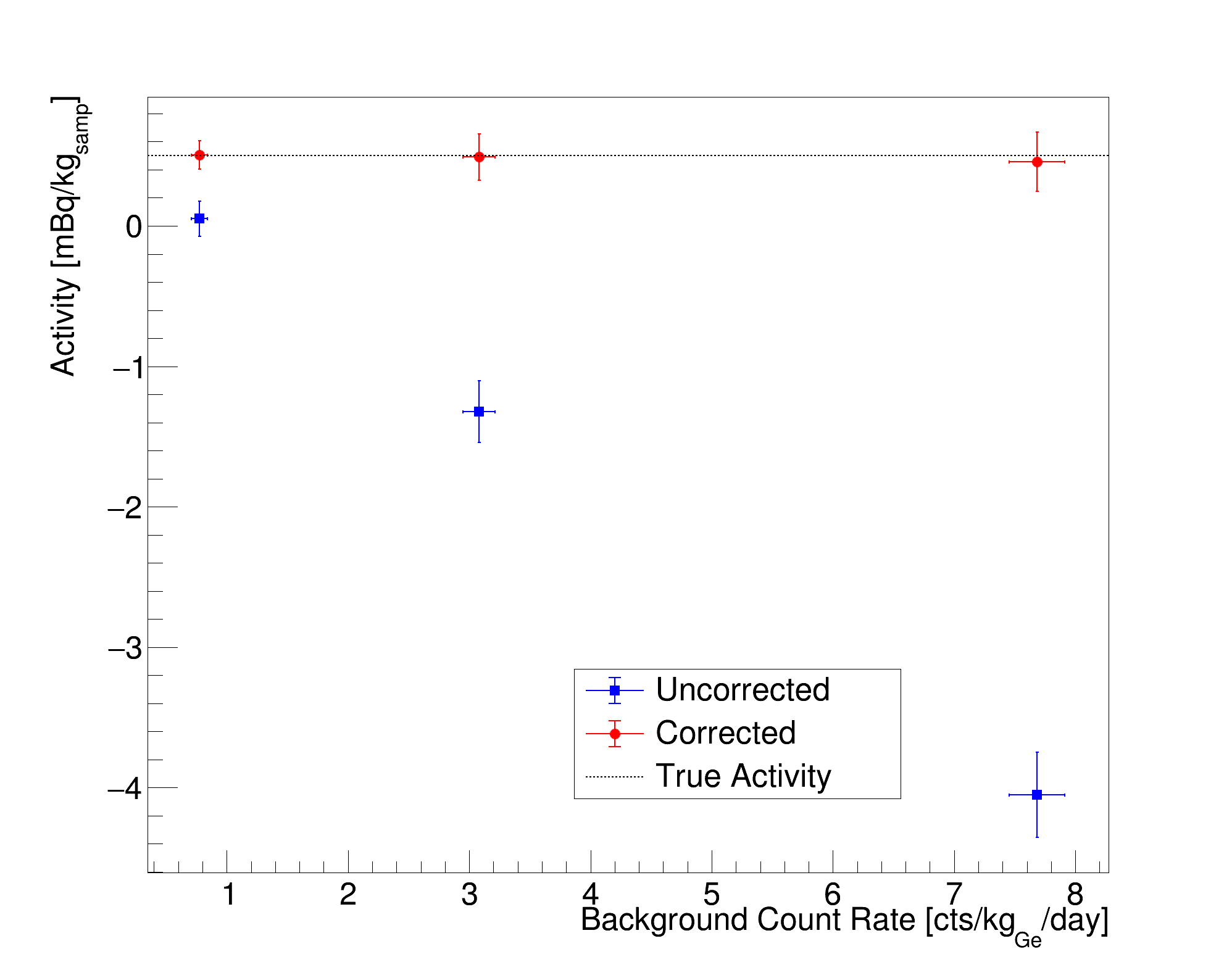}
    \caption{\label{fig:activity_295_external}}
  \end{subfigure}
  \caption{The measured activity for (a) the 609.3~keV characteristic photons from $^{214}$Bi and (b) the 295.2~keV characteristic photons from $^{214}$Pb in the presence of external backgrounds, shown with and without the background shielding correction applied.}
  \label{fig:external}
\end{figure}

The uncorrected activities observed for low, medium, and high background levels are consistently underestimated compared with the true simulated sample activity (Figure~\ref{fig:external}). When the appropriate corrections are applied for the given $E_\gamma$, the apparent sample activity is consistent with the true value.

\subsubsection{Internal Backgrounds}
\label{sec:internalbkg}

In a real detector, some backgrounds can be assumed to be almost entirely internal to the sample. For example, $^{232}$Th is a common contaminant in the metal forging process of steel detector supports and in FETs which comprise the front-end electronics for the detector, which are all found internally.

\begin{figure}
  \centering
  \begin{minipage}{0.48\textwidth}
    \centering
    \includegraphics[width=\textwidth]{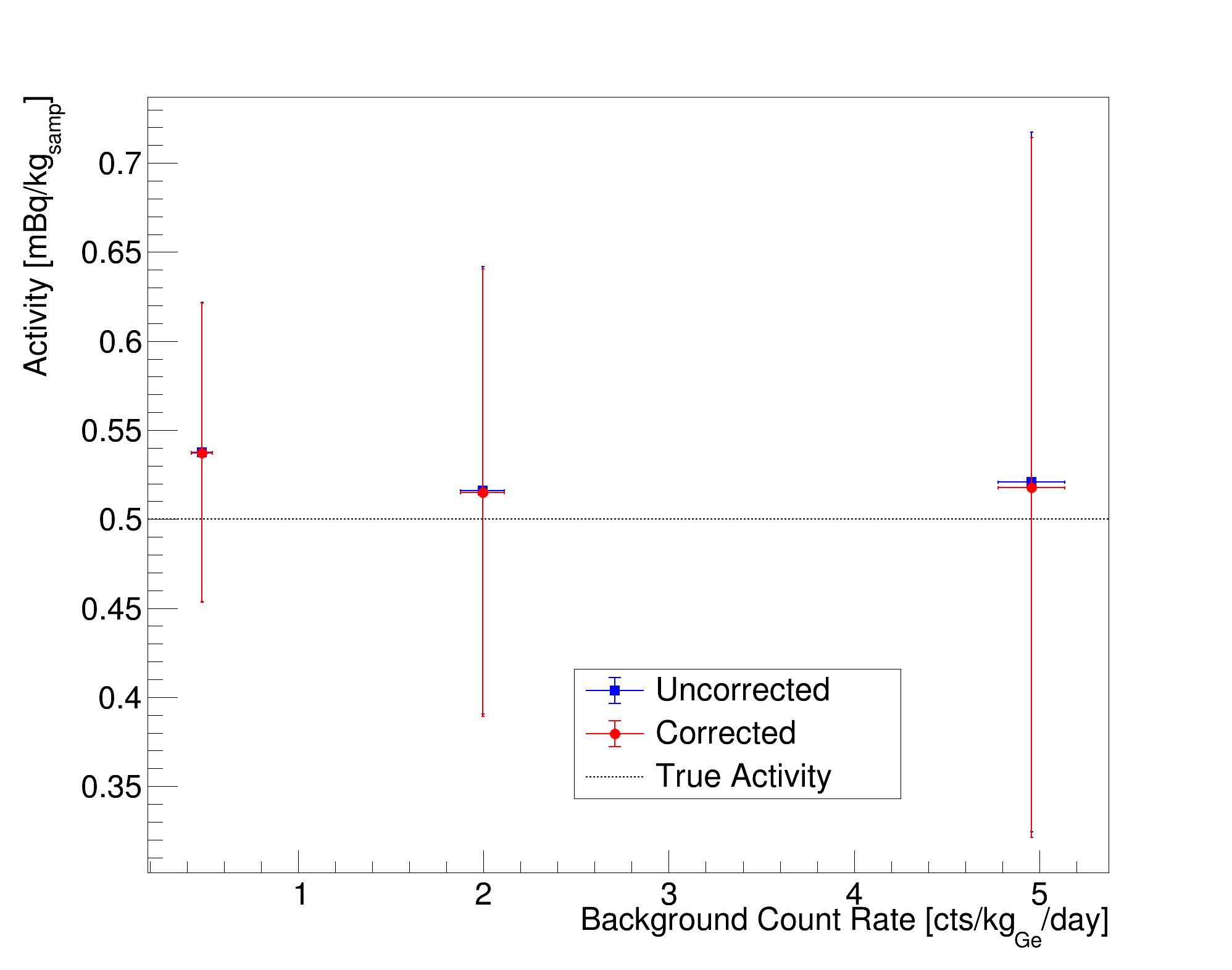}
    \caption{The measured activity for the 583.2~keV characteristic photons from $^{208}$Tl in the presence of internal backgrounds, shown with and without the background shielding correction applied.}
    \label{fig:activity_583_internal}
  \end{minipage}\hfill
  \begin{minipage}{0.48\textwidth}
    \centering
    \includegraphics[width=\textwidth]{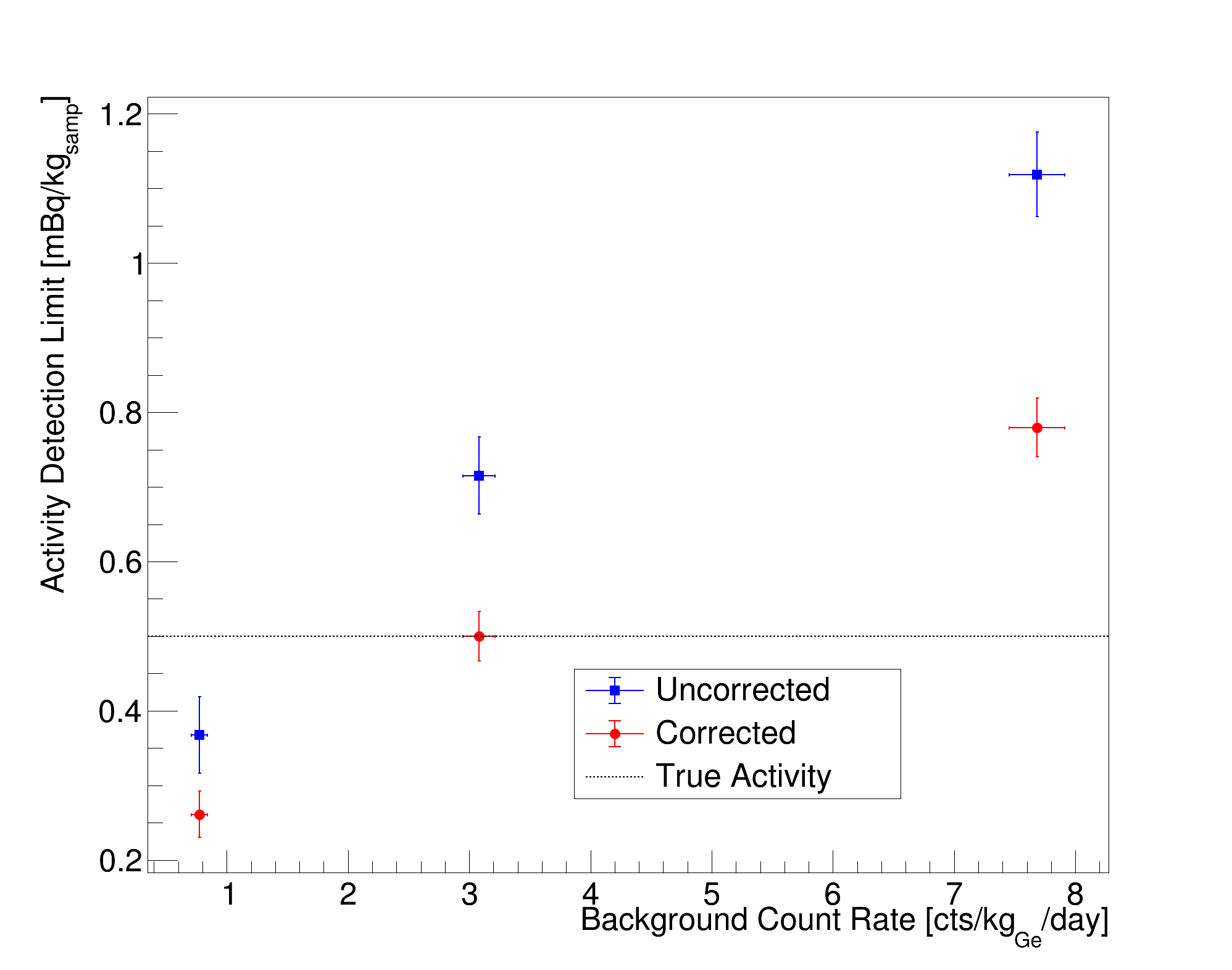}
    \caption{The detector $L_D$ for the 295.2~keV characteristic photons from $^{214}$Pb in the presence of external backgrounds, shown with and without the background shielding correction applied.}
    \label{fig:detlim_295_external}
  \end{minipage}
\end{figure}

The 583.2~keV characteristic photon from $^{208}$Tl in the later part of the $^{232}$Th decay chain is investigated as an internal background. Figure~\ref{fig:activity_583_internal} shows that the measured activity of this characteristic photon is accurate to the true simulated sample activity for internal backgrounds even without applying any background correction. As expected from Figure~\ref{fig:internalab}, purely internal backgrounds are not significantly affected by the background shielding effect. Even below 500~keV, the cumulative effect of the enhancement of $b_I$ would cause only a slight overestimation of the observed activity and a slight underestimation of $L_D$ due to increased backscattering in the Compton continuum.

\subsubsection{Detection Limits}
\label{sec:detlim}

The uncorrected ($a=b=1$) $L_D$ is overestimated compared with the true $L_D$ when the sample is on the detector. When the sample activity is similar to or lower than the background, $L_D$ is improved (reduced) by correcting the total background count rate, $B$. In some cases, the corrected background count rate may allow a statistically significant activity measurement which may not have been possible if uncorrected. The limit of detection for measuring the 295.2~keV photons from external sources of $^{214}$Pb (Figure~\ref{fig:detlim_295_external}) shows the improvement in detector sensitivity achieved by correcting for the background shielding effect.

\subsubsection{Negligible Backgrounds}
\label{sec:nobkg}

When the background of a particular nuclide is negligible because it is not present in the background or because the sample nuclide activity is significantly greater than the background, correcting for the background attenuation by the sample has little effect. In this case, $\mbkgspec\left(E_\gamma\right)_{\mathrm{net}}\ll\msampspec\left(E_\gamma\right)_{\mathrm{net}}$, so $P_S$ in Equation~\ref{eq:netsampcorr} still holds even if the correction factor, $a$, is not implemented or poorly estimated. However, the total background count rate, $B$, must still be corrected as in Equation~\ref{eq:totbkgcorr} because $b$ describes the effect of the sample material on the total background Compton continuum, not just the continuum of the nuclide in question. 

\begin{figure}
  \begin{subfigure}{0.49\textwidth}
    \centering
    \includegraphics[width=\textwidth]{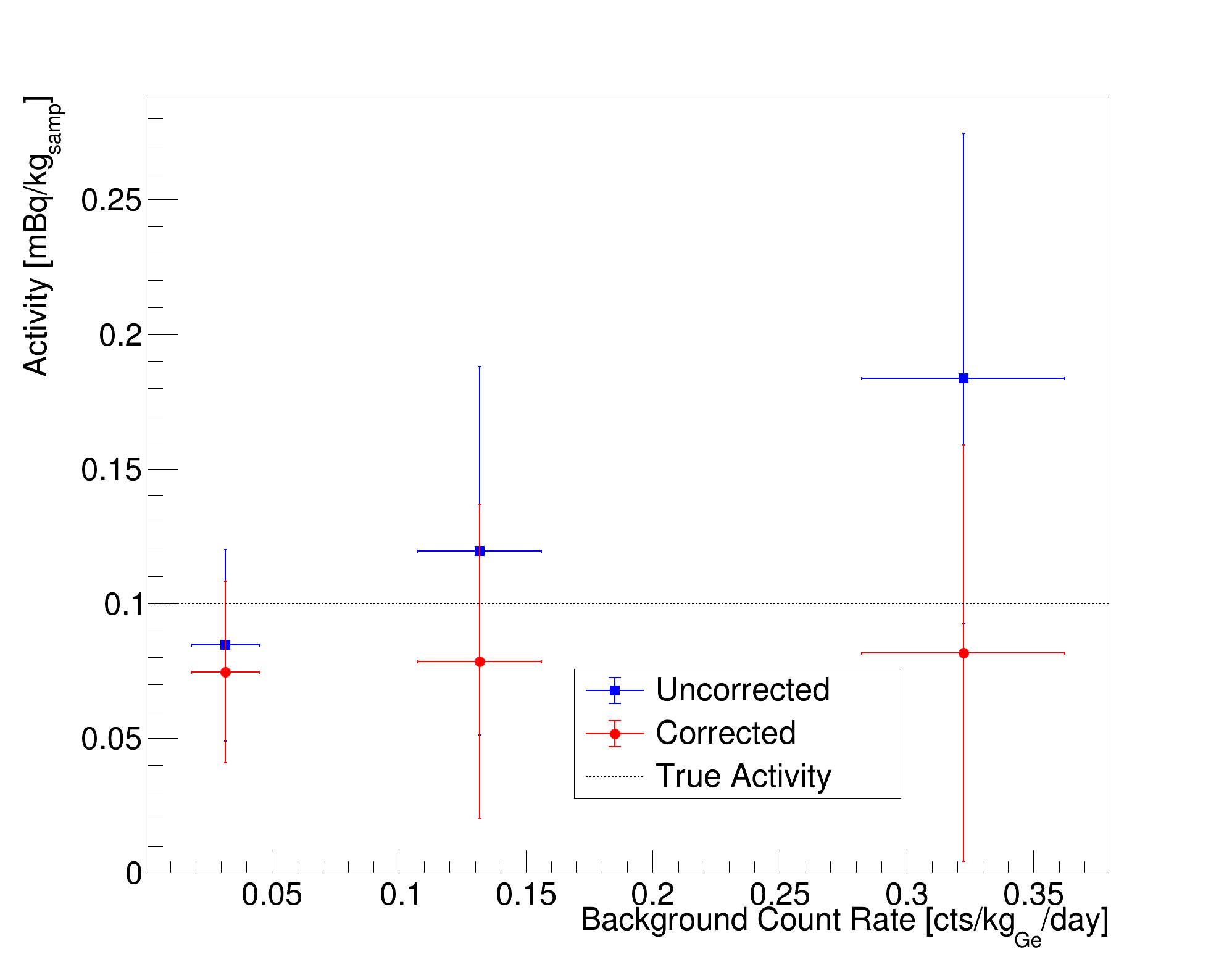}
    \caption{\label{fig:activity_788_external}}
  \end{subfigure}
  \hfill
  \begin{subfigure}{0.49\textwidth}
    \centering
    \includegraphics[width=\textwidth]{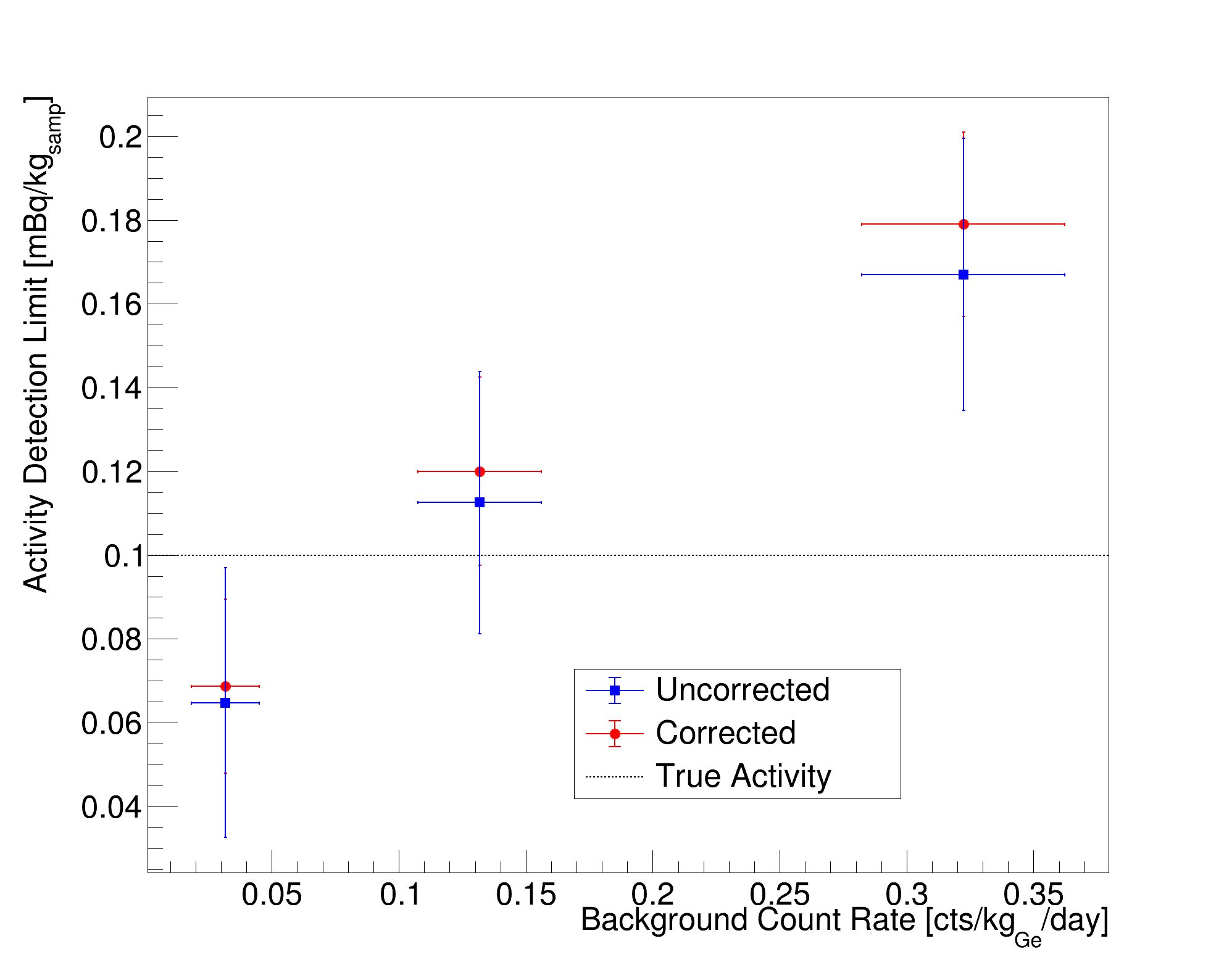}
    \caption{\label{fig:detlim_788_external}}
  \end{subfigure}
  \caption{The (a) measured sample activity and (b) detector $L_D$ for the 788.7~keV characteristic photons in the presence of external backgrounds, shown with and without the background shielding correction applied.}
  \label{fig:788_external}
\end{figure}

The analysis of the 788.7~keV characteristic photons from $^{138}$La in the current simulation is interesting because the nuclide is not present in the background but is present in the sample at an activity of 0.1~mBq/kg. Figure~\ref{fig:788_external} shows slight shifts in the measured sample activity and detector $L_D$ by correcting the sample's effect on the background continuum alone.

\section{Application To A Real Sample Measurement}
\label{sec:realmeas}

For some low-activity \gd\ sample measurements on a low background HPGe detector at Boulby Underground Laboratory, significant negative net count rates are observed for some nuclides following background subtraction. For these nuclides, the true background source distribution is unknown but predicted to be predominantly external to the sample. The measurement, therefore, suffers from systematic errors caused by background shielding by the extensive, dense sample material. The external background sources are assumed to be isotropic in distribution as sources include airborne $^{222}$Rn, impurities in the Cu lining, and impurities in the Pb shield, which are distributed evenly around the HPGe and sample. For these backgrounds, it is predicted that the simplified external background model of Section~\ref{sec:correcting} can appropriately correct the background shielding effect for the problematic nuclide measurements.

\begin{sidewaystable}[tbp]
  \centering
\begin{tabular}{@{}llrrrrrrr@{}}
\toprule
\multirow{2}{*}{Decay Chain} &
  \multirow{2}{*}{Nuclide} &
  \multirow{2}{*}{\begin{tabular}[c]{@{}r@{}}Energy \\ {[keV]}\end{tabular}} &
  \multicolumn{2}{c}{Uncorrected} &
  \multicolumn{2}{c}{External Bkg Corrected} &
  \multicolumn{2}{c}{Internal Bkg Corrected} \\ \cmidrule(lr){4-5}\cmidrule(lr){6-7}\cmidrule(lr){8-9} 
 &
   &
   &
  \begin{tabular}[c]{@{}r@{}}Activity \\ {[mBq/kg]}\end{tabular} &
  \begin{tabular}[c]{@{}r@{}}$L_D$ \\ {[mBq/kg]}\end{tabular} &
  \begin{tabular}[c]{@{}r@{}}Activity\\ {[mBq/kg]}\end{tabular} &
  \begin{tabular}[c]{@{}r@{}}$L_D$\\ {[mBq/kg]}\end{tabular} &
  \begin{tabular}[c]{@{}r@{}}Activity\\ {[mBq/kg]}\end{tabular} &
  \begin{tabular}[c]{@{}r@{}}$L_D$\\ {[mBq/kg]}\end{tabular} \\ \midrule
$^{238}$U (early)  & $^{234m}$Pa & 1001.03 & 4.66$\pm$2.15    & 6.47  & 3.67$\pm$1.91    & 6.33  & 4.65$\pm$2.16    & 6.48  \\ \midrule
$^{238}$U (late)   & $^{214}$Pb  & 351.93  & -1.23$\pm$0.15   & 0.56  & 0.22$\pm$0.07    & 0.41  & -1.22$\pm$0.16   & 0.56  \\
$^{238}$U (late)   & $^{214}$Pb  & 295.22  & -2.08$\pm$0.28   & 0.90  & -0.063$\pm$0.058 & 0.654 & -2.07$\pm$0.29   & 0.91  \\
$^{238}$U (late)   & $^{214}$Bi  & 609.31  & -1.06$\pm$0.14   & 0.52  & -0.16$\pm$0.08   & 0.42  & -1.06$\pm$0.15   & 0.52  \\
$^{238}$U (late)   & $^{214}$Bi  & 1120.29 & -0.29$\pm$0.14   & 1.19  & 0.22$\pm$0.14    & 1.06  & -0.29$\pm$0.15   & 1.19  \\
$^{238}$U (late)   & $^{214}$Bi  & 1764.49 & -1.75$\pm$0.39   & 1.17  & -1.08$\pm$0.33   & 1.04  & -1.76$\pm$0.39   & 1.17  \\ \midrule
$^{232}$Th (early) & $^{228}$Ac  & 338.32  & 0.62$\pm$0.18    & 0.65  & 0.82$\pm$0.21    & 0.61  & 0.62$\pm$0.18    & 0.67  \\
$^{232}$Th (early) & $^{228}$Ac  & 911.20  & 0.28$\pm$0.09    & 0.31  & 0.38$\pm$0.11    & 0.27  & 0.28$\pm$0.09    & 0.31  \\
$^{232}$Th (early) & $^{228}$Ac  & 968.96  & -0.012$\pm$0.025 & 0.497 & 0.085$\pm$0.068  & 0.447 & -0.012$\pm$0.025 & 0.498 \\ \midrule
$^{232}$Th (late)  & $^{212}$Pb  & 238.63  & 0.40$\pm$0.08    & 0.29  & 0.51$\pm$0.09    & 0.27  & 0.40$\pm$0.08    & 0.32  \\
$^{232}$Th (late)  & $^{212}$Bi  & 727.33  & 0.40$\pm$0.21    & 1.07  & 0.42$\pm$0.22    & 1.01  & 0.40$\pm$0.21    & 1.07  \\
$^{232}$Th (late)  & $^{208}$Tl  & 583.19  & 0.32$\pm$0.09    & 0.32  & 0.44$\pm$0.11    & 0.29  & 0.32$\pm$0.09    & 0.32  \\
$^{232}$Th (late)  & $^{208}$Tl  & 2614.51 & -0.23$\pm$0.10   & 0.44  & -0.085$\pm$0.072 & 0.395 & -0.23$\pm$0.10   & 0.44  \\ \midrule
$^{235}$U (early)  & $^{235}$U   & 163.36  & 1.56$\pm$0.56    & 2.90  & -0.18$\pm$0.19   & 2.93  & 1.57$\pm$0.57    & 3.22  \\
$^{235}$U (early)  & $^{235}$U   & 143.77  & 0.43$\pm$0.23    & 2.13  & 1.02$\pm$0.36    & 1.82  & 0.43$\pm$0.23    & 2.28  \\
$^{235}$U (early)  & $^{235}$U   & 185.72  & -0.026$\pm$0.019 & 0.237 & 0.044$\pm$0.025  & 0.210 & -0.026$\pm$0.019 & 0.258 \\ \midrule
$^{235}$U (late)   & $^{227}$Th  & 256.23  & 0.78$\pm$0.28    & 1.25  & 0.82$\pm$0.29    & 1.23  & 0.78$\pm$0.28    & 1.37  \\
$^{235}$U (late)   & $^{227}$Th  & 235.92  & 0.22$\pm$0.10    & 0.71  & -0.31$\pm$0.12   & 0.78  & 0.21$\pm$0.10    & 0.81  \\
$^{235}$U (late)   & $^{223}$Ra  & 269.46  & 1.31$\pm$0.26    & 0.56  & 1.21$\pm$0.25    & 0.58  & 1.30$\pm$0.26    & 0.61  \\
$^{235}$U (late)   & $^{219}$Rn  & 271.23  & 1.10$\pm$0.26    & 0.67  & 0.84$\pm$0.23    & 0.73  & 1.10$\pm$0.26    & 0.74  \\ \midrule
                   & $^{176}$Lu  & 201.83  & 1.91$\pm$0.16    & 0.15  & 1.85$\pm$0.15    & 0.16  & 1.91$\pm$0.16    & 0.17  \\
                   & $^{176}$Lu  & 306.82  & 1.67$\pm$0.12    & 0.09  & 1.67$\pm$0.12    & 0.09  & 1.67$\pm$0.12    & 0.09  \\ \midrule
                   & $^{40}$K    & 1460.82 & -2.57$\pm$0.57   & 2.32  & -0.55$\pm$0.44   & 2.03  & -2.59$\pm$0.60   & 2.32  \\ \midrule
                   & $^{60}$Co   & 1173.23 & -0.14$\pm$0.04   & 0.14  & -0.067$\pm$0.028 & 0.122 & -0.14$\pm$0.04   & 0.14  \\
                   & $^{60}$Co   & 1332.49 & -0.13$\pm$0.04   & 0.16  & -0.053$\pm$0.027 & 0.137 & -0.13$\pm$0.04   & 0.16  \\ \midrule
                   & $^{137}$Cs  & 661.66  & -0.025$\pm$0.015 & 0.085 & -0.014$\pm$0.011 & 0.079 & -0.025$\pm$0.015 & 0.086 \\ \midrule
                   & $^{138}$La  & 788.74  & -0.008$\pm$0.014 & 0.213 & 0.027$\pm$0.026  & 0.193 & -0.008$\pm$0.014 & 0.213 \\
                   & $^{138}$La  & 1435.80 & 0.072$\pm$0.033  & 0.074 & 0.066$\pm$0.032  & 0.070 & 0.072$\pm$0.033  & 0.074 \\ \bottomrule
\end{tabular}
  \caption{List of the measured nuclide activities and $L_D$ for a real \gd\ sample at Boulby Underground Laboratory. Reported activity uncertainties are the combination of counting statistical uncertainty and the known systematic uncertainties, such as errors in full-energy peak efficiency and true coincidence summing. Errors due to the background shielding by the sample material are excluded from the uncertainty estimate for this study. Corrections are made for the background shielding effect of the 5~kg sample under two different assumptions: all backgrounds are external, and all backgrounds are internal.}
  \label{tab:gdsamp}
\end{sidewaystable}

In Table~\ref{tab:gdsamp}, statistically significant negative (around $-5\sigma$ to $-7\sigma$) net count rates are observed for late-chain $^{238}$U, $^{40}$K, and $^{60}$Co, indicating that there must be an external background component of these nuclides which is larger than the sample nuclide activity and the internal background contribution ($\alpha\ll\beta$). Some nuclide activities, like $^{176}$Lu, are significantly greater than $L_D$, and others are consistent with zero. This measurement aims to discover or place limits on the true sample radioactivity, which is unknown.

For the external-only background correction ($\alpha=0$), the significantly negative uncorrected activities are adjusted to be consistent with zero at the $2\sigma$ level. Even though the external-only background model is the most extreme correction applied for isotropically distributed backgrounds, the resulting corrected activity is reasonable. Since it is consistent with zero, the maximally corrected activity is brought to the minimum physically-possible value, indicating that the backgrounds are highly likely to be entirely external to the sample. The remaining slightly negative apparent activity may be caused by deviations of the detector background from the isotropic source location distribution assumed in the calculation of the correction factors or another systematic error such as discrepancies in the geometric modelling of the system in the MC simulation.

Several possible background source distributions may explain the uncorrected activities that are not significantly negative. The dominant background source may be internal to the sample location ($\alpha>\beta$), or the dominant external background contribution may be small ($\alpha<\beta$) and overshadowed by a true non-zero sample activity ($P_S\gg P_B$). In this case, one should consider whether there is evidence of the nuclides in the external detector background. If not, the observed sample activity is already unbiased without applying corrections. However, if the nuclide is present externally, then additional information about the precise location of the background source must be used to accurately correct for the background shielding effect.

In certain scenarios, such as in material screening programmes for rare-event physics detector construction where the goal is non-contamination, it is best to take a conservative approach in reporting the activity of samples if the background source locations cannot be determined. For example, when it is not clear if a background source is internal or external, both extreme corrections may be performed, and the largest of the resulting $L_D$ may be reported. An overestimated $L_D$ can still be overcome by increasing the sample measurement time. In some cases where the net count rate is significantly negative, applying an external-only background correction may be the most sensible approach. Where the nuclide backgrounds are not observed in \tbkgspec\ and a significant sample activity is measured, it may be best not to apply any corrections. The best results will be obtained when the locations of background nuclides are known and can be accurately simulated, or a suitably radiopure standard can be acquired.

\section{Conclusions}
\label{sec:conclusions}

There is potential for a systematic error in HPGe-based gamma spectrometric measurements of low activity samples on low background detectors. A bias occurs when detector backgrounds are shielded by the sample material, which is enhanced when it is large, dense, and covers a large solid angle of the HPGe crystal. Then the no-sample, no-standard background spectrum does not represent the background present in the sample spectrum.

A framework and method of correction for this effect is proposed, which uses MC simulation techniques already widely used by modern HPGe laboratories. The calculation of correction factors depends on the background source distribution in the detector system. If the distribution is known, corrections can be applied that accurately recover the true sample activity. Even when the background source distribution is not known \textit{a priori}, some conservative, reasonable assumptions can be made when applied to a sample measurement with unknown activity. With a complete understanding of the sources of detector backgrounds, the proposed framework and procedure can accurately correct the background shielding effect for real low-activity sample measurements.

\section*{Acknowledgements}

The authors acknowledge the invaluable support of the U.K. Science and Technology Facilities Council (STFC) Boulby Underground Laboratory and Israel Chemicals Ltd UK (ICL-UK) for providing the facilities and access required to perform this work. This work was supported by the STFC under award numbers ST/R000069/1, ST/V002821/1 and ST/V006185/1.

\bibliography{refs}

\end{document}